\documentclass[aps,pra,a4paper,twocolumn,showpacs]{revtex4}
\usepackage[dvips]{epsfig}
\usepackage{amsmath}
\usepackage{amssymb}
\usepackage{subfigure}

\begin{document}
\preprint{JChemPhys}

\title{Quantum  dynamics of the O + OH $\to$ H + O$_2$ reaction at low temperatures}

\author{Goulven Qu{\'e}m{\'e}ner, Naduvalath Balakrishnan}
\affiliation{Department of Chemistry, University of Nevada Las Vegas,
Las Vegas, Nevada 89154, USA}

\author{Brian K. Kendrick}
\affiliation{Theoretical Division, Los Alamos National Laboratory,
Los Alamos, New Mexico 87545, USA}

\date{\today}

\begin{abstract}
We report quantum dynamics calculations of the
O + OH $\to$ H + O$_2$
reaction on two different representations of the
electronic ground state potential energy surface (PES)
using a time-independent
quantum formalism based on hyperspherical coordinates. Calculations show 
that several excited vibrational levels of the product
 O$_2$ molecule are populated in the reaction. Rate coefficients evaluated
using both PESs were found to be very sensitive to the energy resolution of the
reaction probability, especially at temperatures lower than 100~K. 
It is found that the rate coefficient remains largely constant in the temperature
range 10 -- 39~K, in agreement with the conclusions of a recent experimental study
[Carty et al., J. Phys. Chem. A {\bf 110}, 3101 (2006)].
This is in contrast with the time-independent quantum calculations of Xu et al. [J. Chem. Phys. {\bf 127}, 024304 (2007)] which, using the same PES,
predicted  two orders of magnitude drop in the rate coefficient value from 39~K to 10~K.
Implications
of our findings to oxygen chemistry in the interstellar medium are discussed.
\end{abstract}

\maketitle

\font\smallfont=cmr7

\section{Introduction}

The reaction
\begin{eqnarray*}
\text{O$(^3\text{P})$ + OH$(^2\Pi)$ $\to$ H$(^2\text{S})$ + O$_2(^3\Sigma_g^-)$} 
\end{eqnarray*}
is of considerable importance in atmospheric, 
combustion and interstellar chemistry.
In the upper stratosphere and lower mesosphere, it plays a major role in partitioning the relative abundance
of OH and HO$_2$ molecules~\cite{Lee82,Clary84,Summers97}. The reaction also plays a 
key role in the night-time airglow emissions from the hydroxyl radical
which
has also been recently detected in  the atmosphere of Venus~\cite{Piccioni08}.
It has been identified as a key reaction in  
interstellar oxygen chemistry~\cite{Dalgarno76,Wagner87,Graff87}
and it
is considered to be  the most important source of oxygen molecules in cold interstellar 
clouds~\cite{Viti01,Smith03}. 
As a consequence, the HO$_2$ system has 
been 
the topic of numerous 
electronic structure calculations of its 
potential energy surface~\cite{Melius79,Pastrana90,Kendrick95,Troe01,Xu05,Xie07} as well as
dynamics calculations to evaluate its temperature dependent rate 
coefficients~\cite{Miller86,Davidsson88,Graff90,Varandas92,Germann97,Skinner98,Viel98,Harding00,Troe01,Xu05,Xu07,Lin08,Jorfi08}.
The reaction has also been the focus of a large number of experimental 
measurements~\cite{Lewis80,Howard80,Howard81,Cohen83,Shin91,Smith94,Baulch94,Robertson02,Robertson06,Carty06}. However,
there still remains significant discrepancy between measured and computed values of its rate
coefficients for temperatures below 200~K,
 which is the most important region for astrophysical
and upper atmospheric applications.

Recently, Carty et al.~\cite{Carty06} reported an experimental measurement of
rate coefficients for the reaction in the temperature
 range 39~K to 142~K and observed no variation
in the rate coefficients with temperature in this regime. Based on their findings they
concluded that 
 the rate coefficient would largely remain constant between 39~K to 10~K, temperatures typical
of cold interstellar clouds.
In contrast to the experimental results of Carty et al., in a recent theoretical work
using a time-independent quantum mechanical (TIQM) method and a $J$-shifting approximation,
Xu et al.~\cite{Xu07} reported a rate coefficient that precipitously dropped by about
two orders of magnitude between 39~K and 10~K.
Using a
time-dependent wave packet (TDWP) method
that does not employ the $J$-shifting approximation,
Lin et al.~\cite{Lin08}
reported a rate coefficient
at 10~K
that is about
one order of magnitude
smaller than that obtained by Carty et al. at 39~K.
In a subsequent study, Quan et al.~\cite{Quan08}
adopted these rate coefficients for modeling molecular
oxygen chemistry in the interstellar medium.

In this paper,
using an accurate TIQM method and two  different representations
of the electronic ground state of the HO$_2$ system,
we show that
the low temperature rate coefficient of the
O + OH reaction is very sensitive to the dense resonance structures in the energy dependence of
its reaction probability.
By using a very fine
grid of collision energies
in the low and ultralow regime
to compute the reaction probabilities,
we obtain nearly temperature independent
values of the reaction rate coefficients in the range 10 -- 39~K, in agreement
with the conclusions of Carty et al.~\cite{Carty06}.
Our results differ from those  of Xu et al.~\cite{Xu07} which show rapid decrease below 40 K.
However, our results do merge
with those of Xu et al. at temperatures above 300~K where the low energy regime does
not make a significant contribution. We believe that
the  rate coefficient reported here would provide a more accurate value for modeling
oxygen chemistry in the interstellar medium.

A brief description of the computational method is provided in section II followed by 
results and discussion in section III. Conclusions are presented in section IV.

\section{Computational Method}

The calculations have been performed using the
adiabatically adjusting principal-axis hyperspherical (APH) approach
of Pack and Parker~\cite{Pack87}.
In the present work, we employed  two representations of the electronic 
ground state $(1 \ ^2A'')$ of the HO$_2$ system.
We used the 
XXZLG PES calculated by Xu, Xie, Zhang, Lin, and Guo~\cite{Xu05,Xie07} which 
was also employed in the study of Xu et al.~\cite{Xu07}.
For the other surface, we used the diatomics-in-molecule (DIM) PES 
developed by Kendrick and Pack~\cite{Kendrick95}, hereafter referred to as the 
DIMKP PES.
For each value of the total angular momentum quantum number, $J$, 
and each value of the hyperspherical radius, $\rho$,
the wavefunction is expanded onto a basis set of adiabatic functions, which are eigenfunctions 
of a triatomic hyperangular Hamiltonian.
The hyperangular Hamiltonian is diagonalized using an Implicitly Restarted Lanczos algorithm and
a hybrid DVR/FBR primitive basis set~\cite{Kendrick99}.
The time-independent Schr{\"o}dinger equation
yields a set of differential close-coupling equations in $\rho$, which is solved using
the log-derivative matrix propagation method of Johnson~\cite{Johnson73}.
The log-derivative matrix is propagated
to a matching distance 
where asymptotic boundary conditions 
are applied to evaluate the reactance matrix, $K^{J}$, and the scattering matrix, $S^{J}$. 
The square elements of the $S^{J}$ matrix provide
the state-to-state transition probabilities, $P^{J}$. 
The matching distance $\rho_m$ and the
number of adiabatic functions $n$ used in the basis set
are determined by optimization and
extensive convergence studies.
The convergence of the $J=0$ reaction
probability $P^{r,J=0}_{v=0,j=0}(E_c)$
for the O + OH$(v=0,j=0)$ reaction
with respect to $\rho_m$ and $n$
is summarized in Table~\ref{TAB1} for the DIMKP PES 
and Table~\ref{TAB2} for the XXZLG PES.
Based on these convergence studies,
we have used $\rho_m=26.8 \ a_0$
and $n=393$ in the final production calculations.

\begin{table}[h]
\begin{center}
\begin{tabular}{c c c c c c}
\hline
$\rho_m$ ($a_0$) & 26.8 & 32.7 & 160.6 & 160.6 & 160.6   \\
$n$ & 393 & 393 & 393 & 380 & 300  \\ [0.5ex]
\hline
$E_c$ (eV) & $P^{r,J=0}_{v=0,j=0}$ & & & & \\
0.001 & 0.7269 & 0.7295 & 0.7284 & 0.7284 & 0.7282   \\
0.01 & 0.3498 & 0.3492 & 0.3490 & 0.3490 & 0.3484  \\
0.1 & 0.2956 & 0.2956 & 0.2954  & 0.2954 & 0.2956   \\
1 & 0.1831 & 0.1841 & 0.1825 & 0.1835 & 0.2653   \\ [1ex]
\hline
\end{tabular}
\end{center}
\caption{
$J=0$ reaction probability of O + OH$(v=0,j=0)$ $\to$ H + O$_2$ for
the DIMKP PES for different values of the matching distance $\rho_m$
and the number of hyperspherical channels $n$ at different collision energies $E_c$.
\label{TAB1}
}
\end{table}

\begin{table}[h]
\begin{center}
\begin{tabular}{c c c c c c}
\hline
$\rho_m$ ($a_0$) & 26.8 & 32.7 & 160.6 & 160.6 & 160.6   \\
$n$ & 393 & 393 & 393 & 380 & 300    \\ [0.5ex]
\hline
$E_c$ (eV) & $P^{r,J=0}_{v=0,j=0}$ & & & & \\
0.001 & 0.2153  & 0.2007 & 0.2081 & 0.2081 & 0.2076   \\
0.01 & 0.2563  & 0.2570 & 0.2577 & 0.2577 & 0.2568  \\
0.1 & 0.1816 & 0.1807 & 0.1795 & 0.1795 & 0.1792   \\
1 & 0.0846 & 0.0846 & 0.0841 & 0.0846 & 0.1355   \\ [1ex]
\hline
\end{tabular}
\end{center}
\caption{
Same as Table~\ref{TAB1} but for the XXZLG PES.
\label{TAB2}
}
\end{table}

\section{Results and discussion}

\subsection{Probabilties}

\begin{figure} [b]
\begin{center}
\includegraphics*[width=8cm,keepaspectratio=true,angle=0]{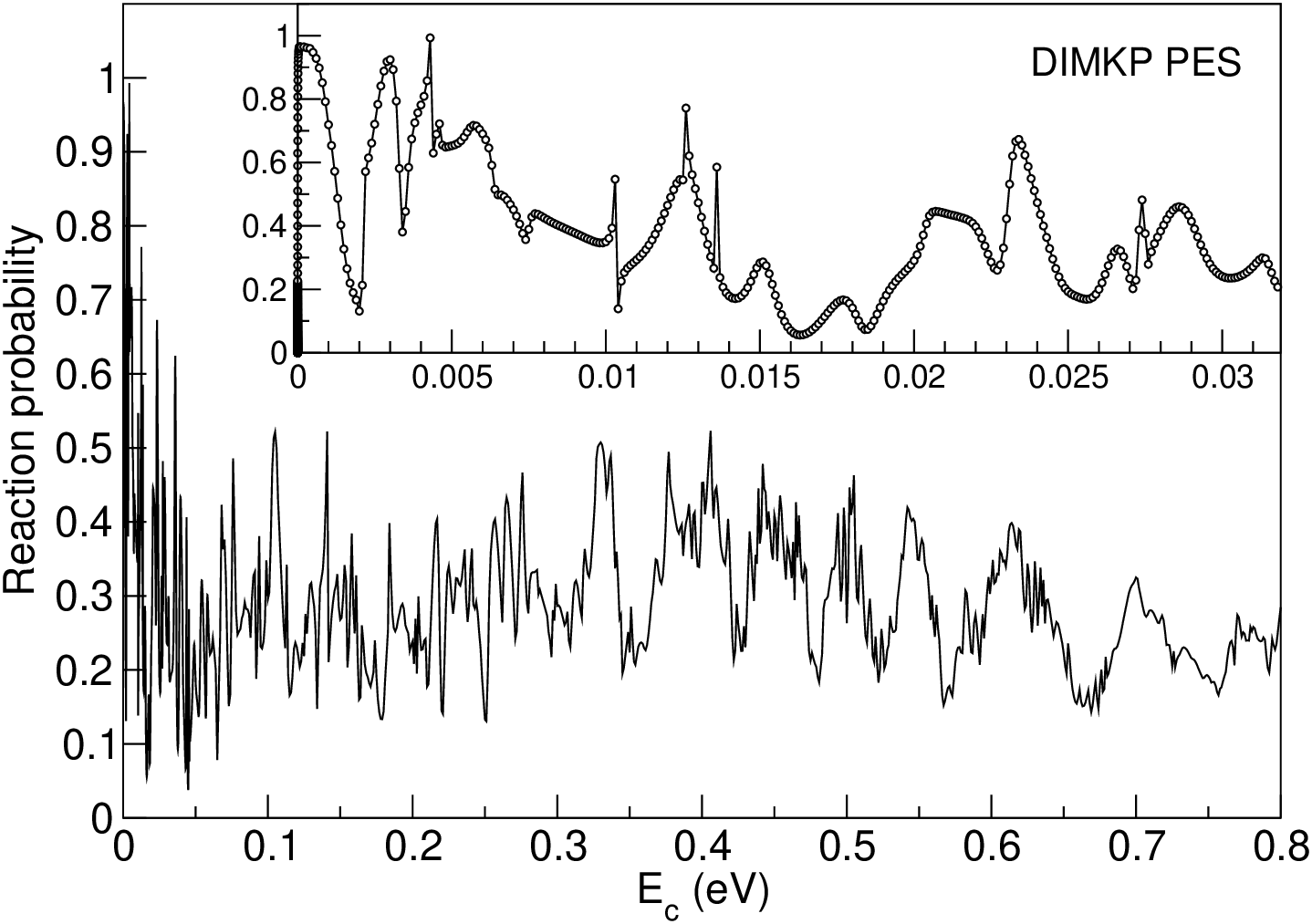} \\
\includegraphics*[width=8cm,keepaspectratio=true,angle=0]{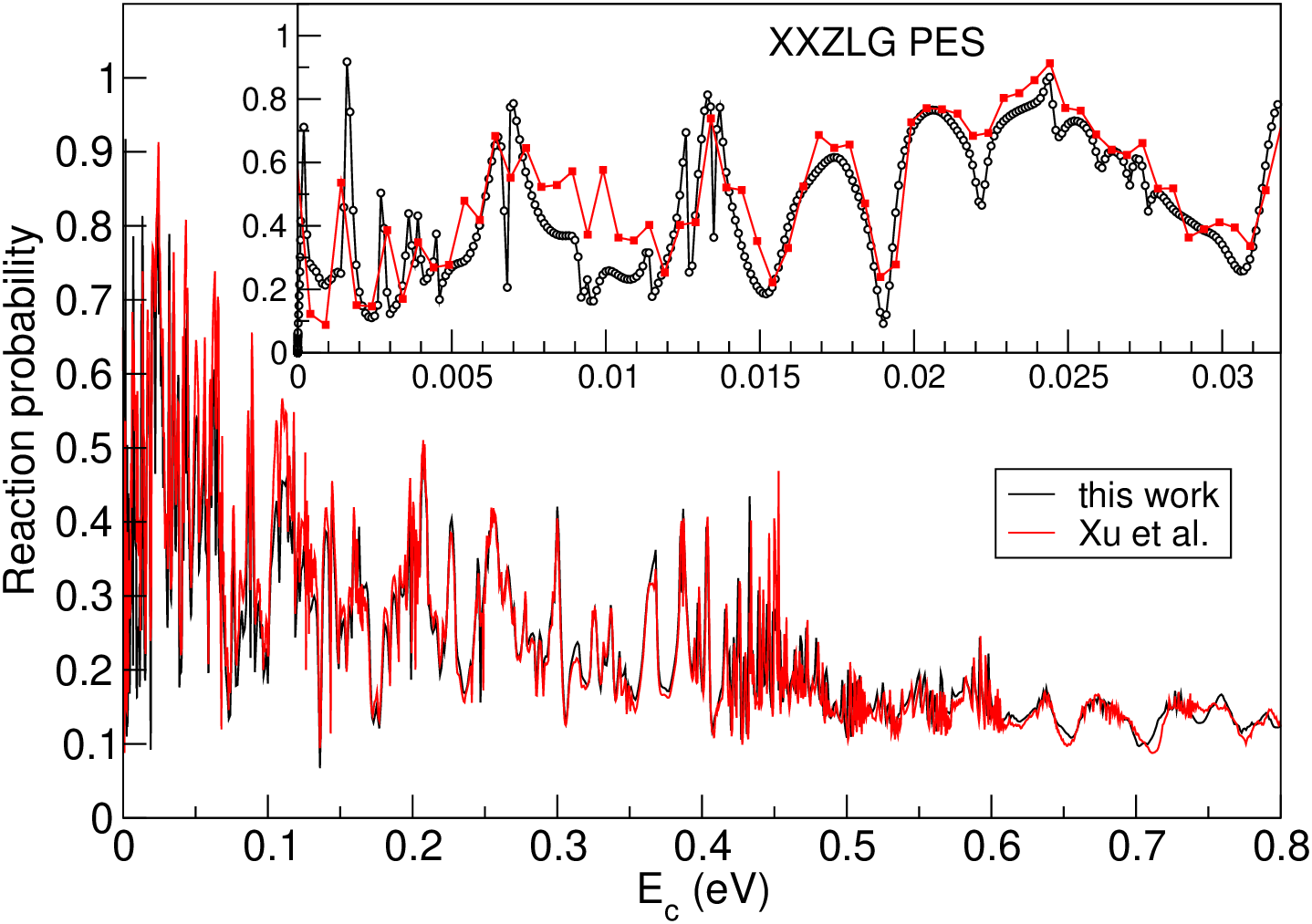}
\caption{
(Color online)
$J=0$ reaction probabilities of
O + OH$(v=0,j=0)$ $\to$ H + O$_2$ reaction on
the DIMKP PES (upper panel)
and the XXZLG PES (lower panel). As discussed in~\cite{Comment1} results of Xu et al.~\cite{Xu07} on the XXZLG PES
have been shifted back by 0.00409~eV to make a one-to-one comparison with our results.
\label{REACPROB-FIG}
}
\end{center}
\end{figure}

\begin{figure}[t]
\begin{center}
\vspace{1cm}
\includegraphics*[width=8cm,angle=0]{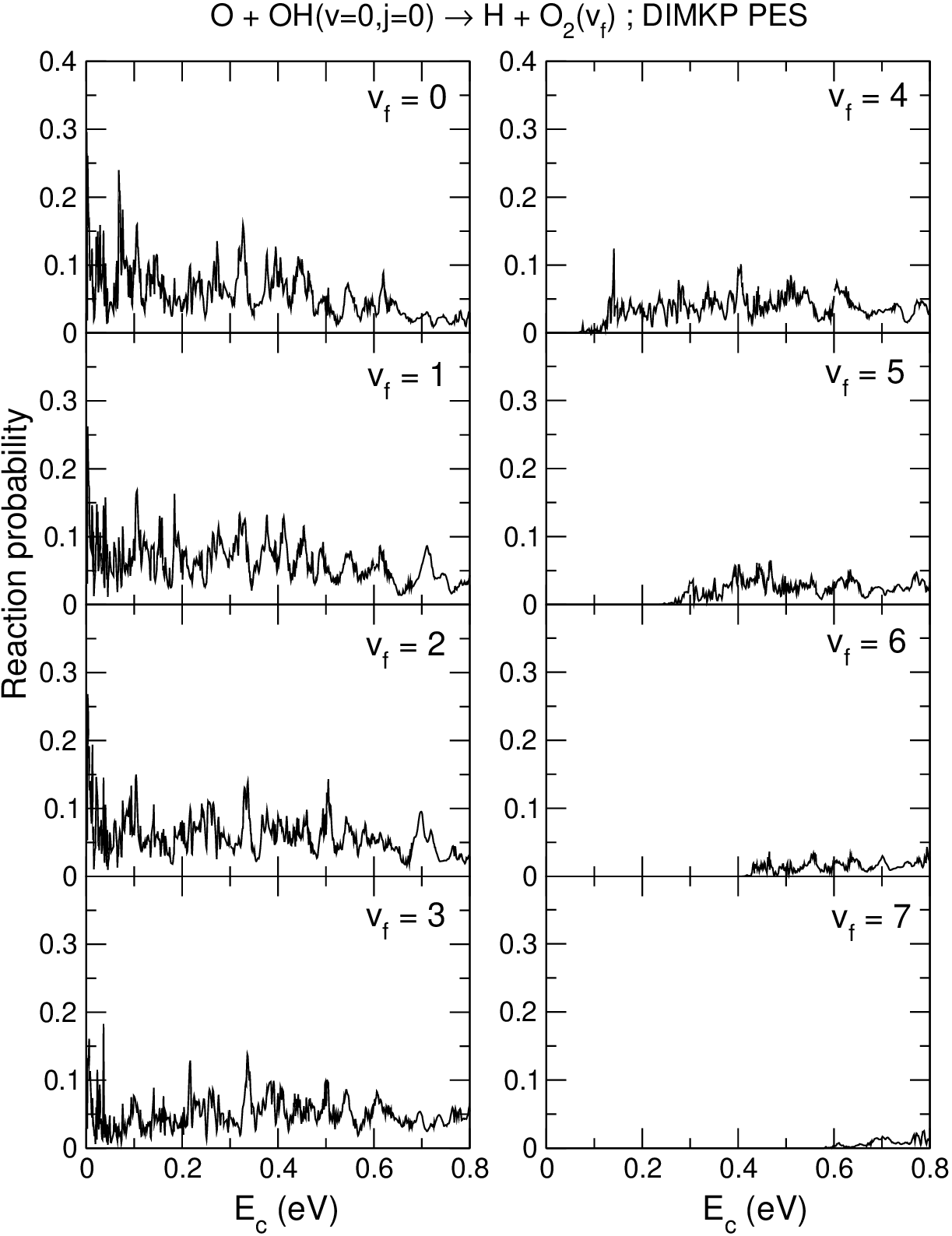}
\caption{
Final vibrational distribution of the $J=0$ reaction probability
of O + OH$(v=0,j=0) \to$ H + O$_2(v_f)$ for the DIMKP PES.
\label{DISTVIB-DIMKP-FIG}
}
\end{center}
\end{figure}

In Fig.~\ref{REACPROB-FIG} we show the $J=0$ reaction probabilities  
as a function of the collision energy 
computed using the DIMKP PES (upper panel) and  the XXZLG PES (lower panel). 
The reaction probabilities calculated by Xu et al. using the XXZLG PES
and a TIQM method are also included in the lower panel.
In both figures, the inset shows 
results for $E_c \le 0.03$~eV to illustrate the energy resolution required for
resolving the resonance features in the reaction probability.
Our calculations include reaction probabilities in the energy range $E_c=[10^{-7} - 0.8]$~eV.
In the energy range $E_c=[10^{-7} - 10^{-4}]$~eV 
where no resonance features are present we included
30 energies in a logarithmic scale. 
Above 0.0001~eV, we used the following energy grids:
$E_c=[0.0001 - 0.0600 ; 0.0001]$~eV and $E_c=[0.060 - 0.800 ; 0.001]$~eV
using linear intervals where the last number in the brackets indicates the energy spacing.
The reaction probabilities obtained using the 
two PESs are not identical and they illustrate the sensitivity of results
to details of the interaction potential.
The global mean is of about 0.3 for the DIMKP PES while it is 
about 0.2 for the XXZLG PES.
The overall trend of the reaction probabilities 
for the XXZLG PES is a decrease with increase in the collision energy
while it oscillates around a value of about 0.3
for the DIMKP PES.
The large number of resonances 
seen in the reaction probabilities
comes from quasi-bound states of the HO$_2$ complex.
We note that our results on the XXZLG PES are in excellent agreement
with those of Xu et al. (red curves)
at high and intermediate collision energies.
However,
some discrepancies appear for low energies $E_c < 0.012$~eV
as seen in the inset of the figure. As discussed in the next
section, these discrepancies at the lower energies can lead to
significantly different rate coefficients at low
temperatures.

\begin{figure}[t]
\begin{center}
\vspace{1cm}
\includegraphics*[width=8cm,angle=0]{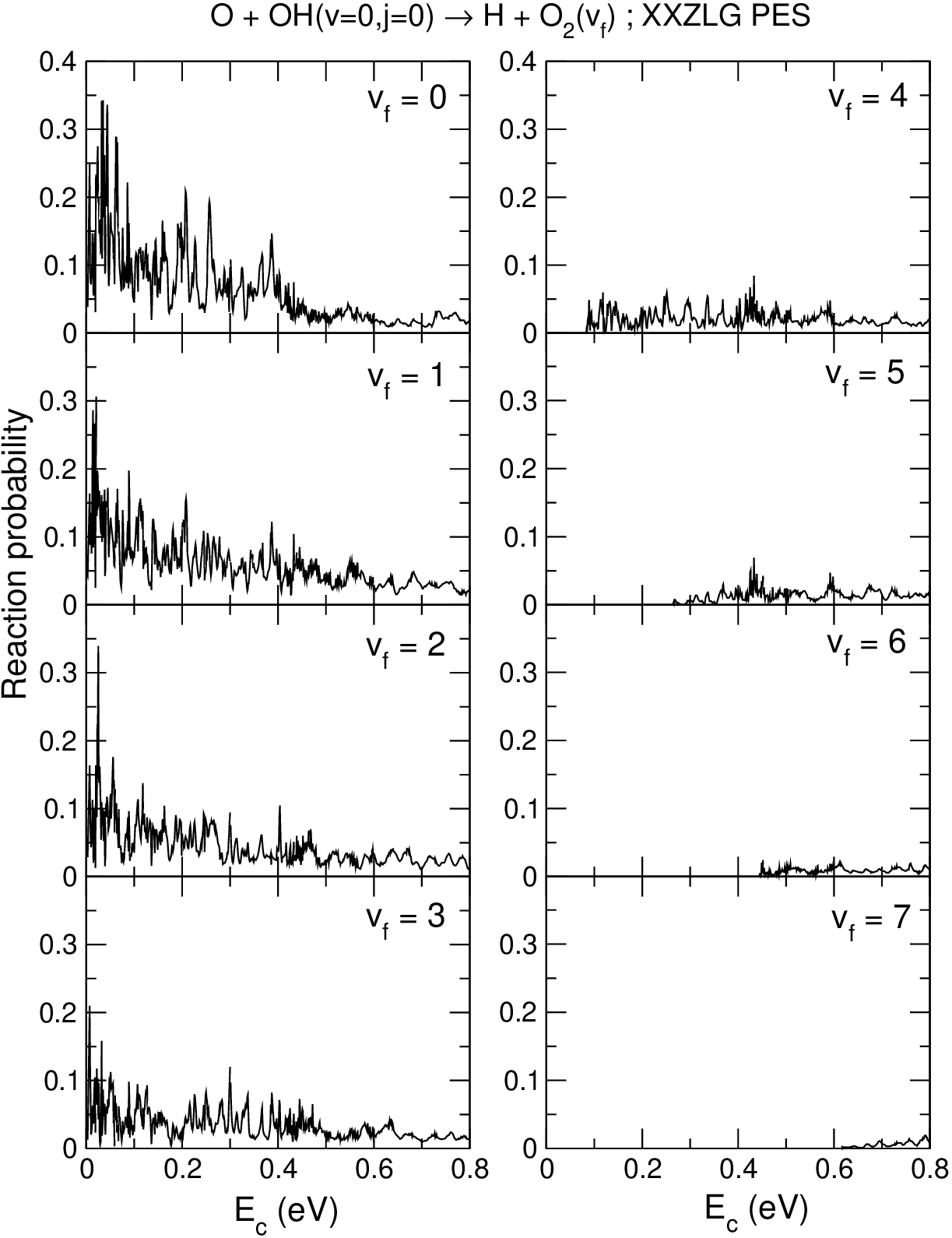}
\caption{
Same as Fig.~\ref{DISTVIB-DIMKP-FIG} but for
the XXZLG PES.
\label{DISTVIB-XXZLG-FIG}
}
\end{center}
\end{figure}

Since the O + OH$(v=0,j=0)$ $\to$ H + O$_2(v_f)$ reaction is exoergic,
the reaction populates a number of excited vibrational levels of the O$_2$ molecule. 
The relative kinetic energy of the ensuing products depends on the vibrational population
of the O$_2$ molecule. Since  non-equilibirium kinetics is an important issue 
in upper atmospheric and astrophysical environments nascent
 vibrational populations of the O$_2$ 
molecules resulting from the O + OH reaction will be important in modeling hydroxyl and
oxygen chemistry in these environments.
The computed vibrational distributions are shown  
in Fig.~\ref{DISTVIB-DIMKP-FIG} for the DIMKP PES
and in Fig.~\ref{DISTVIB-XXZLG-FIG} for the XXZLG PES.
The results for the DIMKP PES show 
that for collision energies $E_c < 0.4$~eV,
 O$_2$ molecules are formed preferentially 
in low-lying vibrational levels $v_f=0-3$. These levels 
are 
open for the O + OH reaction, even in the limit of 
vanishing collision energies.
For  collision energies $E_c > 0.4$~eV,
formation of O$_2$ molecules
in $v_f=4,5$ also competes with the lower vibrational levels.
The results for the XXZLG PES in Fig.~\ref{DISTVIB-XXZLG-FIG}
show
similar trends.
Vibrational levels $v_f=0-3$ 
are the most probable for $E_c < 0.4$~eV,
while  $v_f=4,5$ are equally probable for $E_c > 0.4$~eV.
For $v_f=0-3$ and $E_c < 0.4$~eV, the
reaction probabilities on the XXZLG PES
are larger than that of the DIMKP PES. The opposite is true 
for $E_c > 0.4$~eV.
Overall, the reaction probabilities for a given $v_f$ are less sensitive to
the collision energy  for the DIMKP PES
than that of the XXZLG PES, consistent with the  result for
the total reaction probabilities shown
in Fig.~\ref{REACPROB-FIG}.

\subsection{Rate coefficients}

\begin{figure}[h]
\begin{center}
\vspace{1cm}
\includegraphics*[width=8cm,angle=0]{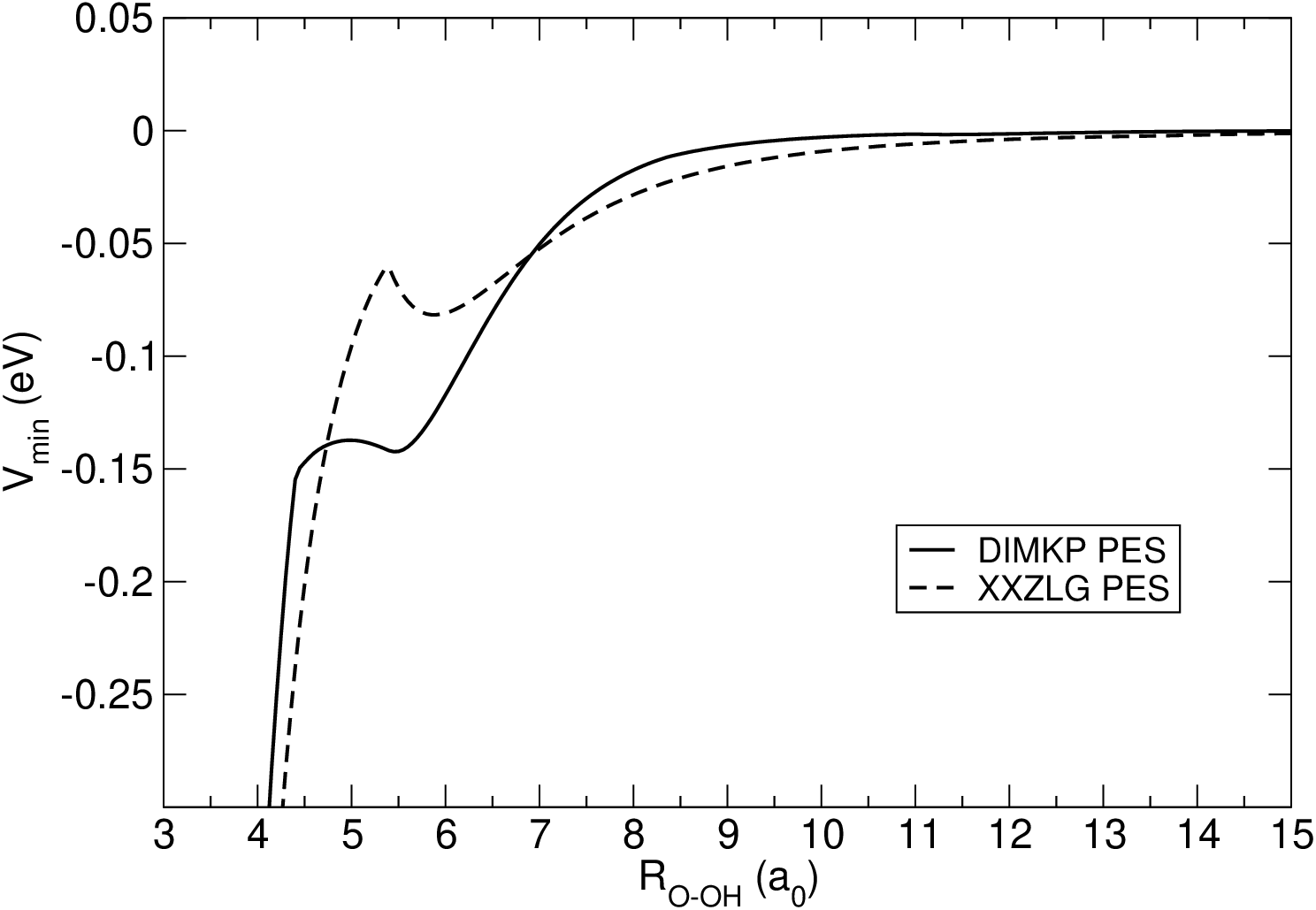}
\caption{
Minimum energy path of the O + OH($v=0,j=0$) reaction as a function of
$R_{\text{O--OH}}$ for the DIMKP and XXZLG PESs.
\label{Minimum-path}
}
\end{center}
\end{figure}

Accurate determination of the rate coefficients would require 
calculations of the reaction probabilities for all contributing values of $J$. 
Computational expense escalates quickly with $J$
unless some angular momentum decoupling approximations are used. 
Like Xu et al., we use the $J$-shifting
approximation~\cite{Bowman91} to compute the initial state-selected rate coefficients for
the reaction.
Within the $J$-shifting approximation, the  rate coefficient is given by the expression:
\begin{multline}
k_{v,j}(T)=\frac{1}{2 \pi \hbar Q_{\text{R}}}
\times \left( \sum_{J}^{} (2J+1) e^{-E^J_{\text{shift}} / (k_B T)}  \right) \\  
\times \int_0^{\infty} P^{r,J=0}_{v,j}(E_c) \ e^{-E_c / (k_B T)} \ dE_c
\label{ratespecified2}
\end{multline}
where $k_B$ is the Boltzmann constant, $P^{r,J=0}_{v,j}$ is
the reaction probability and  $E^J_{\text{shift}}$ 
is the height of the effective barrier for a given partial wave $J$
in the entrance channel. The barrier height is evaluated from the effective
potential, $V_{\text{eff}}^J$, for a given partial wave:
\begin{eqnarray}
V_{\text{eff}}^J=\frac{\hbar^2 J (J+1)}{2 \mu (R_{\text{O--OH}})^2}
+ V_{\text{min}}(R_{\text{O--OH}})
\label{ejshift}
\end{eqnarray}
where $V_{\text{min}}(R_{\text{O--OH}})$
is the minimum energy path of the PES
as a function of $R_{\text{O--OH}}$ and
$\mu$ is the O$-$OH reduced mass. The minimum energy paths for both 
PESs are  shown in Fig. \ref{Minimum-path}. In Eq.~\eqref{ratespecified2},
$Q_{\text{R}} = Q_{\text{trans}} \times  Q_{\text{el}}$ is the reactant partition function. 
For the translational partition function we used the standard formula,
$Q_{\text{trans}}=\left( \frac{\mu k_B T}{2 \pi \hbar^2} \right)^{3/2}$.
For the electronic partition function
we used the expression given by Graff and Wagner~\cite{Graff90}:
\begin{eqnarray*}
Q_{\text{el}} = \frac{(5 + 3 e^{-228/T} + e^{-326/T}) ( 2 + 2 e^{-205/T} )}{2}.
\end{eqnarray*}

\subsubsection{Sensitivity of rate coefficients to energy resolution of reaction 
probabilities }

\begin{figure} [t]
\begin{center}
\includegraphics*[width=8cm,keepaspectratio=true,angle=0]{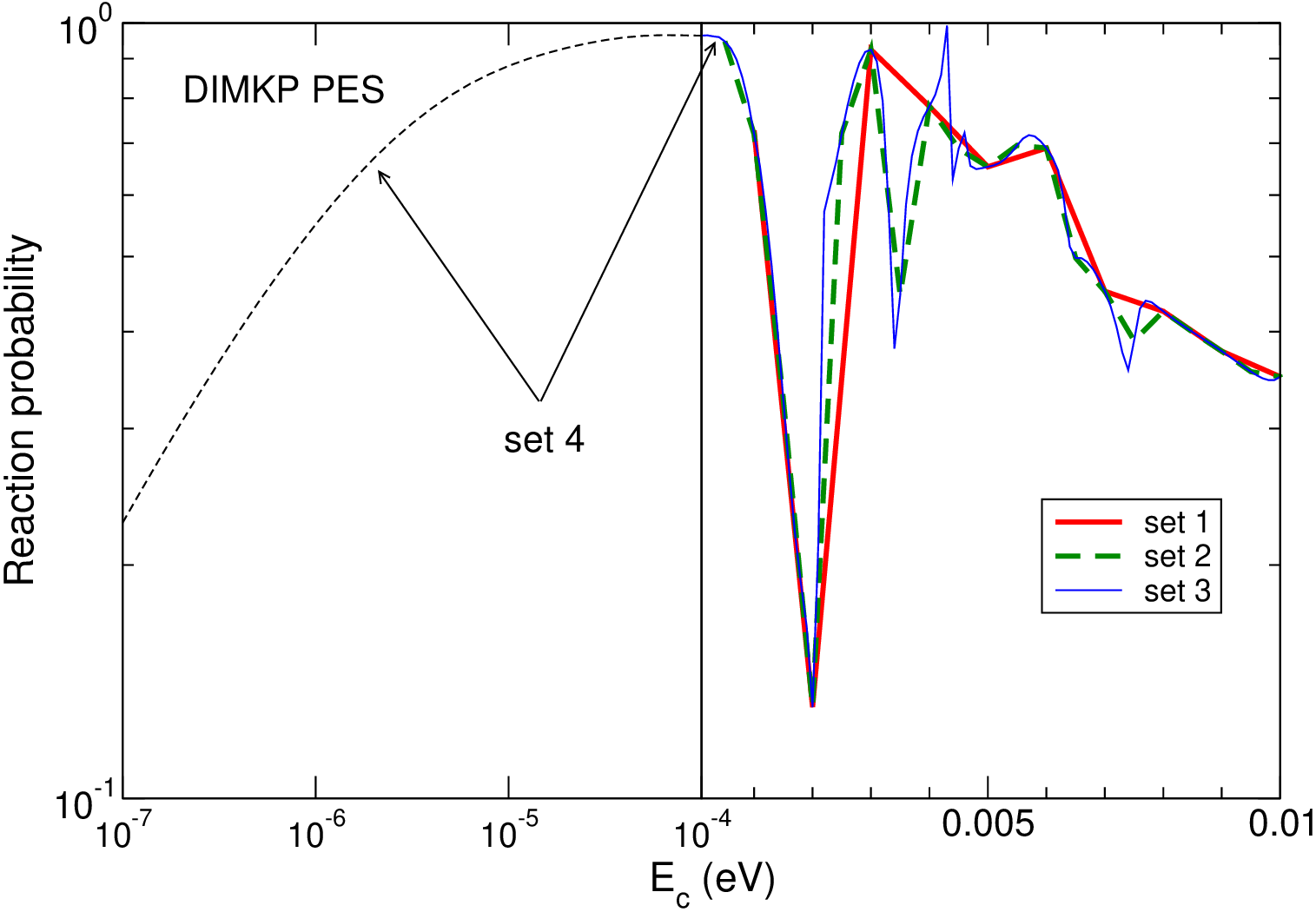} \\
\includegraphics*[width=8cm,keepaspectratio=true,angle=0]{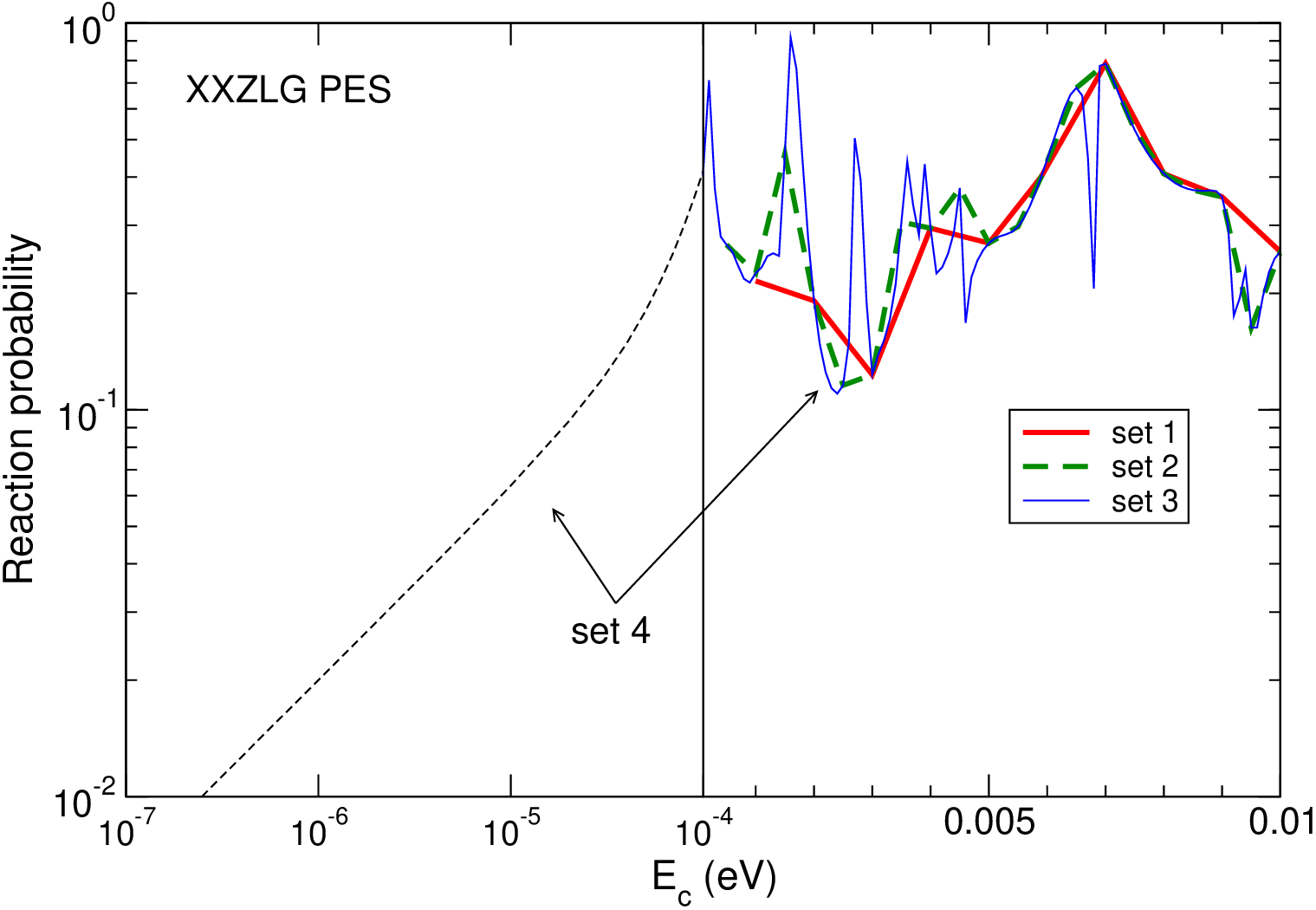}
\caption{
(Color online)
Convergence of the reaction probabilities
as functions of the collision energy
for the DIMKP PES (upper panel)
and for the XXZLG PES (lower panel).
See text for the definition of the sets employed.
\label{PROBREAC-SET-FIG}
}
\end{center}
\end{figure}

\begin{figure} [t]
\begin{center}
\includegraphics*[width=8cm,keepaspectratio=true,angle=0]
{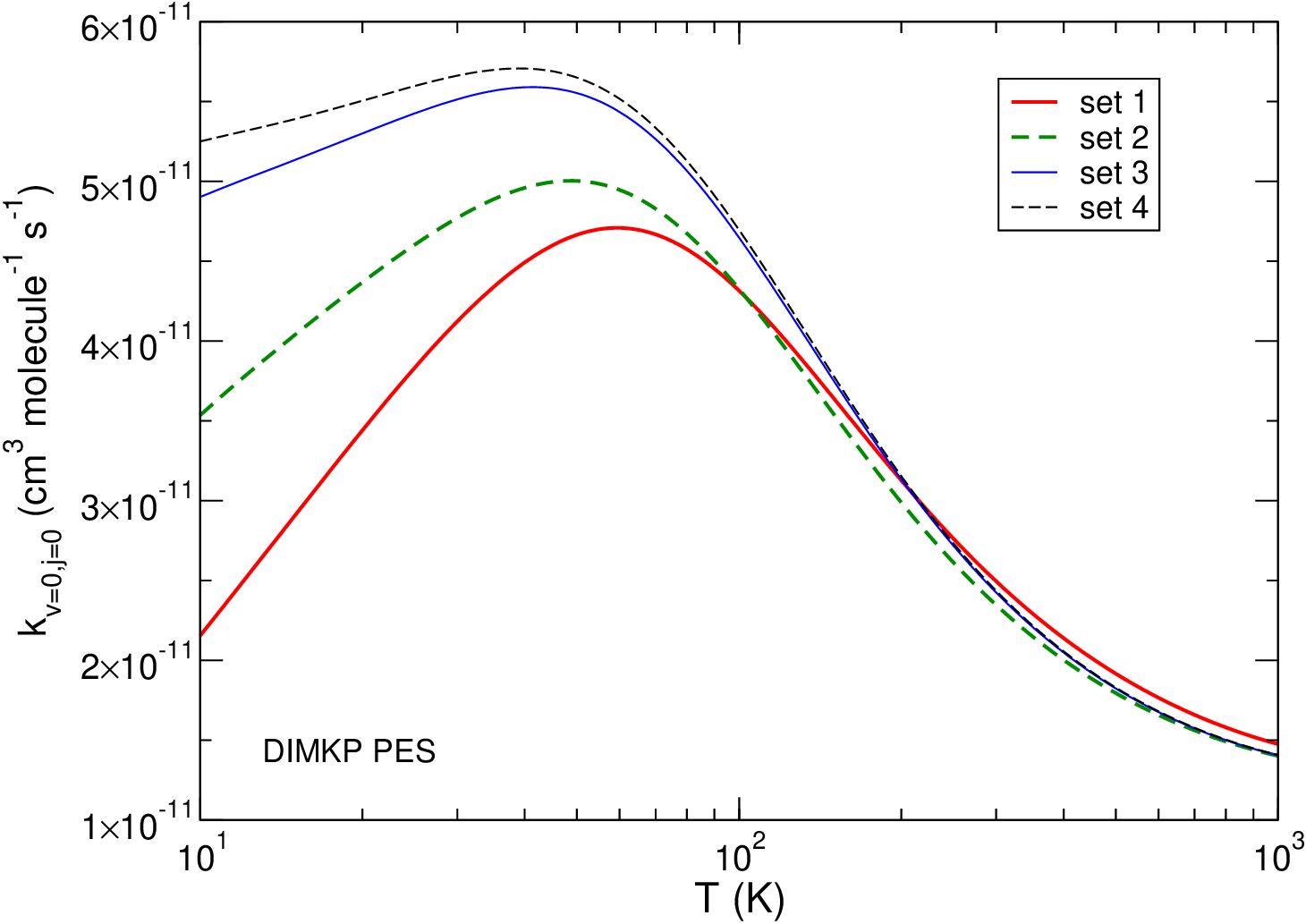} \\
\includegraphics*[width=8cm,keepaspectratio=true,angle=0]
{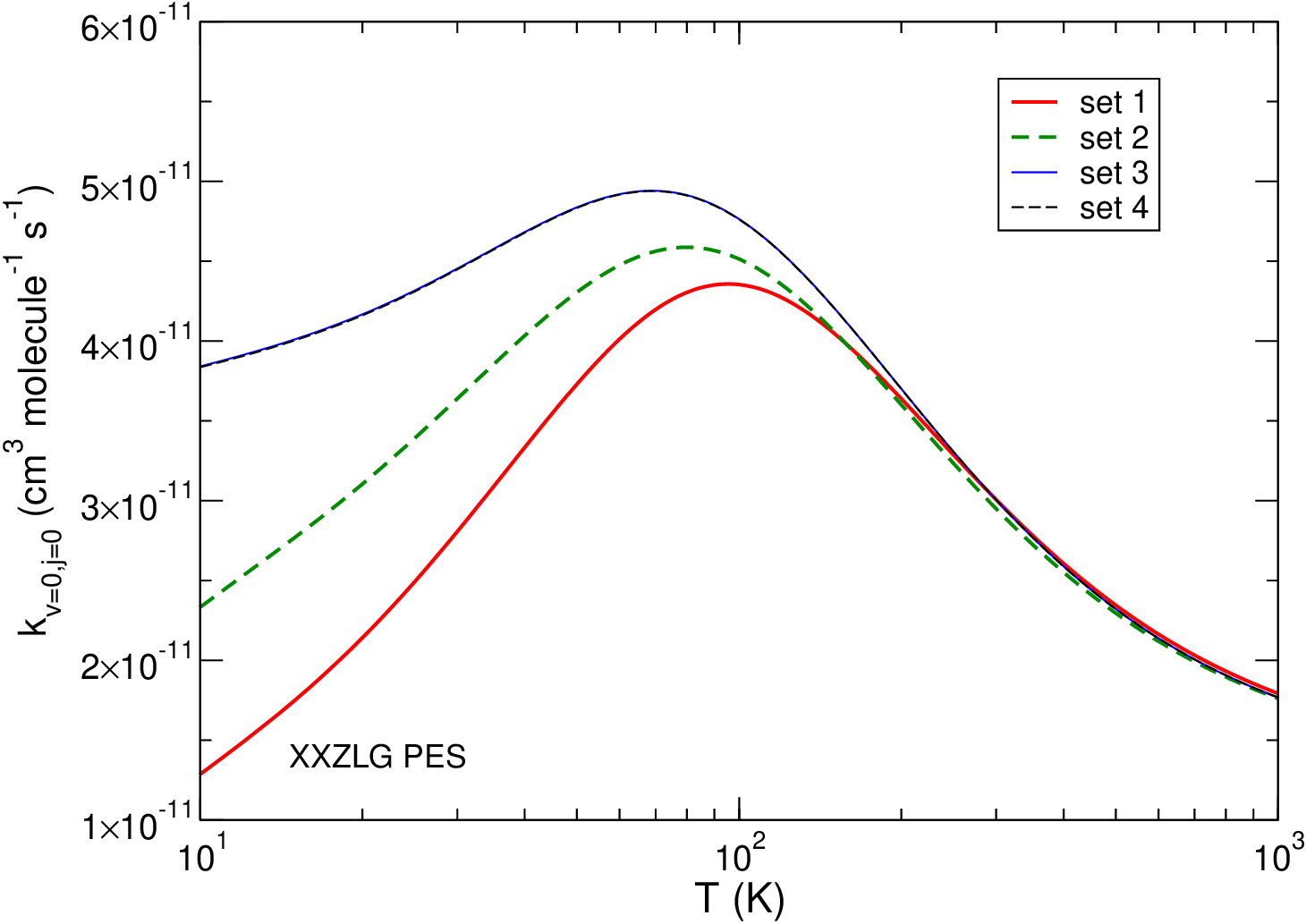}
\caption{
(Color online)
Same as Fig.~\ref{PROBREAC-SET-FIG}
but for the rate coefficients
as functions of the temperature.
\label{RATE-SET-FIG}
}
\end{center}
\end{figure}

Calculation of rate coefficients involves integration 
of the Boltzmann distribution 
weighted by the reaction probability
over the collision energy.
The integration has to be performed carefully
at low temperatures where the Boltzmann distribution can extend to low energies.
This is especially important for capture-type reactions as the present system for
which the rate coefficients
are generally large at low temperatures.
The typical temperature in the cold interstellar clouds
is about 10~K. At this temperature, one needs to include 
collision energies 
lower than $10^{-4}$~eV ($\approx 1$~K)
in accurately calculating the rate coefficients.
It is known  from Bethe--Wigner laws~\cite{Bethe35,Wigner48}
that rate coefficients of
exothermic reactions are finite
in the limit of zero temperatures~\cite{Weck06}.
The limiting value can be large for tunneling dominated 
reactions such as the F + H$_2$ system~\cite{Weck06,Bala01} as well as 
 barrierless reactions such
as  Li + Li$_2(v)$ collisions~\cite{Quemener07}.
Furthermore,  numerous triatomic
resonances can appear
in the probabilities
at collision energies near the reaction threshold which can affect the
value of the rate coefficient at low temperatures.

We have paid careful attention to the convergence
of the reaction probabilities and  rate coefficients 
with the resolution of collision energy grid
at very low energies.
Four sets of collision energies
have been used 
to check the convergence of the rate coefficients. 
Fig.~\ref{PROBREAC-SET-FIG} shows the low-energy portion of
the reaction probabilities employed in the four sets
for the two PESs.
 Using the same notation
as before and with the energies in eV,
the first set
(set 1, bold red curve) corresponds to
$E_c=[0.001 - 0.800 ; 0.001]$.
The second set 
(set 2, dashed green curve) includes
$E_c=[0.0001 - 0.0600 ; 0.0005]$ and $E_c=[0.060 - 0.800 ; 0.001]$.
The third set (set 3, thin blue curve) uses a finer energy resolution in
the low energy regime:  
$E_c=[0.0001 - 0.0600 ; 0.0001]; E_c=[0.060 - 0.800 ; 0.001]$.
The fourth set (set 4, dashed black curve) is composed of set 3 plus the ultralow
energy regime, 
$E_c=[10^{-7} - 10^{-4}]$. 
There is no 
resonance features in the ultralow energy regime where
 the probabilities vary as the square root of the energy,
in accordance with the Bethe--Wigner laws~\cite{Bethe35,Wigner48}. 

As Fig.~\ref{PROBREAC-SET-FIG} shows the ultralow regime
is more prominent for the DIMKP PES due to
the different description of the
long-range interaction potential.
It is seen that the resonances are not fully resolved
in calculations with set 1 and set 2 for both PESs. This 
is especially the case for the XXZLG PES
in the energy range $0.001-0.005$~eV.
Thus, the third multiplicative term
in Eq.~\eqref{ratespecified2} is more accurately evaluated using
set 3 and set 4.

Figure~\ref{RATE-SET-FIG} shows the convergence of the rate coefficients with the
energy resolution for the four collision energy sets for the DIMKP PES (upper panel)
and  the XXZLG PES (lower panel). 
The figure clearly illustrates  that
the rate coefficient at temperatures below 100~K 
 are very sensitive to the resolution of the
energy grid used.  The convergence improves with increase in  energy resolution of the reaction probabilities
and also when the ultralow energy regime is included. The contribution
of the ultralow energy regime is more important for the DIMKP PES as evident from
the corresponding reaction probabilities shown in Fig.~\ref{PROBREAC-SET-FIG}.
Figure~\ref{RATE-SET-FIG} shows
that careful attention must be devoted to the resolution
of the energy grid
and/or to the inclusion of the very low energy regime
in order to accurately calculate the low temperature rate coefficients
of the O + OH reaction. This applies regardless of whether or not a $J$-shifting approximation
is used in the calculation of the rate coefficients.

\subsubsection{Comparison with experiments and other theoretical results}

\begin{figure} [h]
\begin{center}
\includegraphics*[width=8cm,keepaspectratio=true,angle=0]{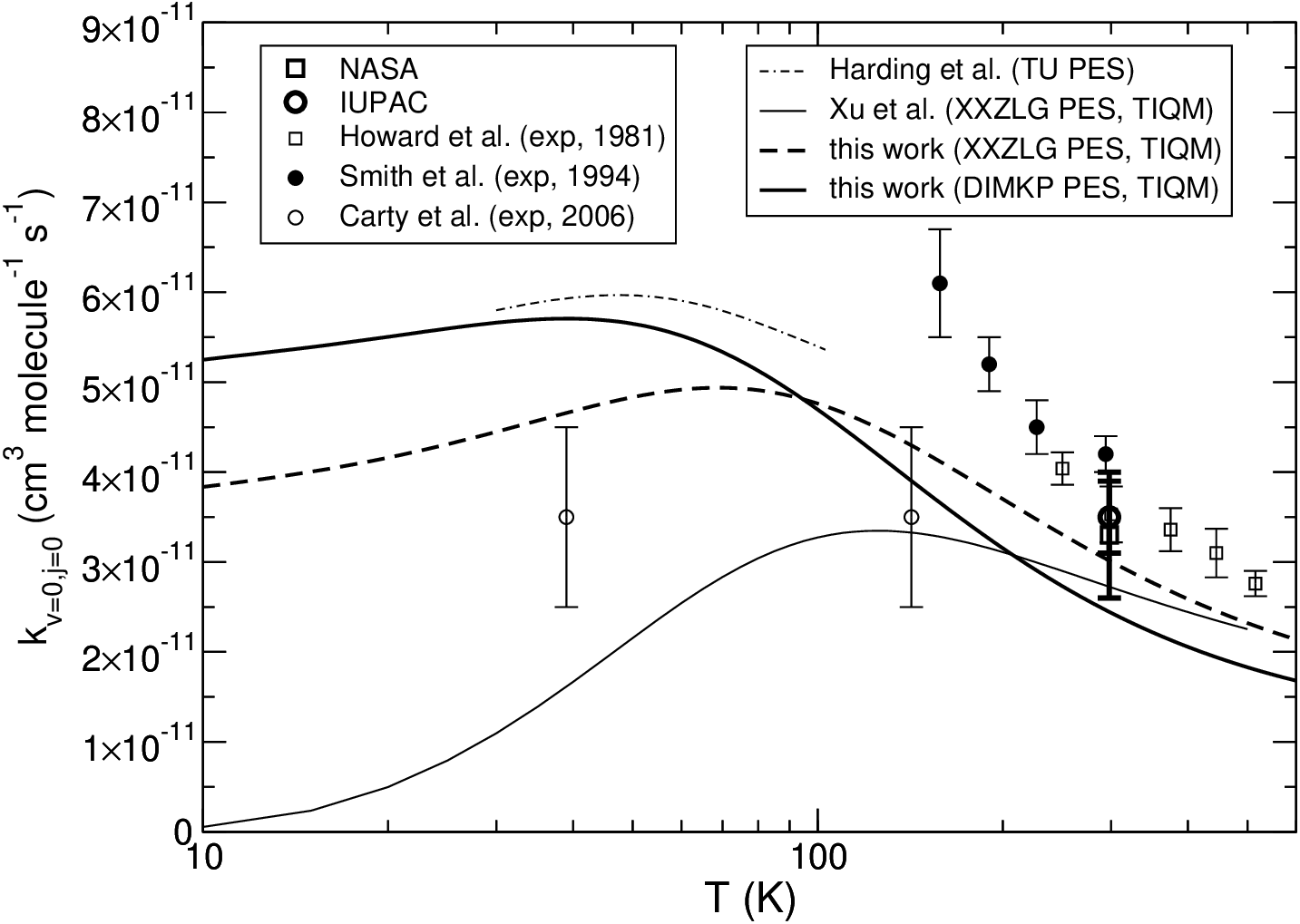}
\caption{
Rate coefficients
of O + OH$(v=0,j=0)$ $\to$ H + O$_2$ as functions of the temperature.
\label{RATE-FIG}
}
\end{center}
\end{figure}

The converged rate coefficients (results from set 4 above) for the $v=0,j=0$ initial
state obtained from the
DIMKP and the XXZLG PESs are shown in Fig.~\ref{RATE-FIG}
as functions of the temperature along with experimental data from
several groups~\cite{Howard81,Smith94,Carty06}
 as well as the recommended values by NASA~\cite{NASA}
and IUPAC~\cite{IUPAC}. The
theoretical results of Xu et al.~\cite{Xu07} obtained using the XXZLG PES
and the results of Harding et al.~\cite{Harding00} obtained
using the PES of Troe and Ushakov~\cite{Troe01} are also included for comparison.
Table~\ref{TAB} lists rate coefficients
at selected temperatures
from different theoretical and experimental studies.

Fig.~\ref{RATE-FIG} shows that the XXZLG PES yields
 results in somewhat better agreement with the experiments 
than the DIMKP PES.
At $T=298$~K, the rate coefficient calculated using  the XXZLG PES
is  3.01$\times 10^{-11}$~cm$^3$~molecule$^{-1}$~s$^{-1}$ compared
to 2.45$\times 10^{-11}$~cm$^3$~molecule$^{-1}$~s$^{-1}$ obtained using the DIMKP PES.
The result on the XXZLG PES
is within the reported error bars of the NASA panel recommended value of
3.3$\pm$0.7$\times 10^{-11}$~cm$^3$~molecule$^{-1}$~s$^{-1}$ at 298~K.
The computed results on the DIMKP and XXZLG PESs
lie within the quoted error bars 
of Carty et al.
at $T=142$~K. 
At $T=39$~K, the result of the XXZLG is slightly above 
the experimental result, while it is higher for the DIMKP PES.

The temperature dependence of the 
rate coefficients predicted by the two PESs is quite similar. 
They
predict rate coefficients within about 30\%
for $10 < T < 100$~K with the DIMKP PES
yielding higher values.
This can be explained by the energy dependence of the
reaction probabilities shown
in the insets of Fig.~\ref{REACPROB-FIG}.
The reaction probabilities are somewhat higher
for the DIMKP PES than for the XXZLG PES
between 0 -- 0.01~eV ($\approx 0 - 100$~K), leading to an increase in
the third multiplicative term in Eq.~\eqref{ratespecified2}.
This can also be explained by the topology of the PESs. As shown in Fig. \ref{Minimum-path}, 
the minimum energy path, $V_{\text{min}}(R_{\text{O--OH}})$,
of the DIMKP PES features a small ``reef"
at $R_{\text{O--OH}}=5.0 \ a_0$. It is located at  $R_{\text{O--OH}}=5.4 \ a_0$
for the XXZLG PES. For the latter PES the reef is about 0.08~eV 
higher than that of the DIMKP PES.
The location and height of the reef is sensitive to the electronic structure method employed.
For certain values of $J$, it becomes an effective barrier.
The smaller reef for the DIMKP PES leads to smaller effective barriers
especially for higher $J$ values.
This enhances the second multiplicative term in Eq.~\eqref{ratespecified2}
and leads
to a higher rate coefficient 
for the DIMKP PES.
For $T>$100~K, 
the two PESs predict rate coefficient
within 20\%
with the DIMKP PES yielding smaller values.
This is attributed
to the smaller overall 
reaction probabilities
of the DIMKP PES compared to the XXZLG PES
for $E_c > 0.01$~eV ($\approx 100$~K),
as seen in the insets
of Fig.~\ref{REACPROB-FIG}.

Our computed rate coefficients do not
show a significant decrease  between 142~K and 39~K, consistent with 
the experimental results of Carty et al.~\cite{Carty06}. Though no experimental
data are available for temperatures below 39~K, our  results on both PESs do not predict 
a dramatic decrease  between 39~K and 10~K.
At 10~K we obtain a rate coefficient of 3.91$\times 10^{-11}$~cm$^3$~molecule$^{-1}$~s$^{-1}$ 
on the XXZLG PES and
5.25$\times 10^{-11}$~cm$^3$~molecule$^{-1}$~s$^{-1}$ on the DIMKP PES.
Our results are about a factor of 70 larger than 
the value of 5.41$\times 10^{-13}$~cm$^3$~molecule$^{-1}$~s$^{-1}$
reported by  Xu et al.~\cite{Xu07}
using the XXZLG PES
and the same $J$-shifting approximation. The discrepancy is attributed to the sparse 
energy grid, small differences in the reaction probabilities for $E_c < 0.012$~eV, 
and the artificial energy shift in the calculations of
Xu et al.~\cite{Comment1,Comment2}.
Among these, the energy shift is the main source of the discrepancy.
Based on our results, we believe that the rate coefficients calculated by
Xu et al. may not be appropriate for modeling oxygen chemistry in the 
interstellar medium~\cite{Quan08}.
While an accurate calculation of
the rate coefficient would require the inclusion of
many higher angular momentum quantum numbers and  non-adiabatic couplings,
the present study shows that special attention must be given
to the energy grid in the calculation of low temperature rate coefficients for capture reactions.

\begin{table}[h]
\begin{center}
\begin{tabular}{c c c c c}
\hline
reference & 10~K & 39~K & 142~K & 298~K  \\ [0.5ex]
\hline
this work, DIMKP & 5.25 & 5.71  & 3.90 & 2.45 \\
this work, XXZLG & 3.91 & 4.66  & 4.30 & 3.01 \\
Xu et al.~\cite{Xu07}, XXZLG & 0.0541  &  & & \\
Lin et al.~\cite{Lin08}, XXZLG & 0.784   &  & & \\
Carty et al.~\cite{Carty06} &  & 3.5$\pm$1.0  & 3.5$\pm$1.0 & \\
NASA~\cite{NASA} &  &   & & 3.3$\pm$0.7 \\
IUPAC~\cite{IUPAC} &   & & & 3.5$\pm$0.4 \\  [1ex]
\hline
\end{tabular}
\end{center}
\caption{
Rate coefficients
of O + OH $\to$ H + O$_2$
in units of
$10^{-11}$~cm$^3$~molecule$^{-1}$~s$^{-1}$
for different temperatures.
\label{TAB}
}
\end{table}

\section{Conclusion}

In conclusion, 
we have performed quantum  dynamics
of the O + OH $\to$ H + O$_2$ reaction
over a wide range of collision energies
on two recent PESs
using a time-independent quantum formalism based on hyperspherical coordinates. We report
total and product vibrational state-selected reaction probabilities, and initial state-selected
rate coefficients for the reaction.
Special attention has been devoted to the convergence of the rate 
coefficients with respect to the energy resolution of reaction probabilities and 
the inclusion of the ultralow energy regime.
The computed rate coefficients 
are in reasonable agreement with experimental measurements
and existing recommended values,
despite using a $J$-shifting approximation for the evaluation of
the rate coefficients. Our calculations show that omission of the 
low energy regime or insufficient energy resolution of the reaction probabilities can lead
to significant errors in the computed rate coefficients. The rate coefficient at 10 K
differ by a factor of 2-3 between calculations using a finer energy grid and a sparse 
energy grid. 
Therefore, we expect that a careful choice of the energy grid
will also be important for an accurate evaluation of the rate coefficients
without the $J$-shifting approximation. 
Overall,
results on the XXZLG PES are in better agreement with the experimental results 
of Carty et al.~\cite{Carty06}
compared to the DIMKP PES in the temperature range of 39$-$142~K. However, only two data
points are available from measurements in this temperature range and
experimental error bars are also quite large.
For both PESs, the  rate coefficients decrease only 
slightly
as the temperature is decreased from 39~K
to 10~K, in agreement with the conclusions of Carty et al.
Based 
on the present results we believe that a re-evaluation of the importance of the O + OH reaction
on O$_2$ abundance would be required for describing oxygen chemistry in the
interstellar medium.

\section{Acknowledgments}

This work was supported  by NSF grants \#PHY-0555565 (N.B.) and \#ATM-0635715 (N.B.).
B.K.K. acknowledges that part of this work was done under the auspices of the US
Department of Energy at Los Alamos National Laboratory. Los Alamos National
Laboratory is operated by Los Alamos National Security,
LLC, for the National Nuclear Security Administration of the
US Department of Energy under contract DE-AC52-
06NA25396. 
We thank D. Xie for providing us the XXZLG PES,
and P. Honvault for providing us the 
reaction probabilities and the rate coefficient
obtained by Xu et al., and for helpful discussions.

\end{document}